\documentclass[a4paper,12pt]{article}
\usepackage[utf8]{inputenc}
\usepackage{graphicx}
\graphicspath{{figures/}}
 \usepackage{caption}
\usepackage{mathrsfs}
\usepackage{amssymb}
\usepackage{textcomp}
\usepackage{amsmath, float}
 \usepackage{times}
\usepackage{color}
\usepackage{enumitem}
\setlist{noitemsep}
\usepackage[left=2cm, right=2cm, top=2cm]{geometry}
\title{\textbf{ Transition Redshift: New constraints from parametric and nonparametric methods}}
\author{Nisha Rani\color{blue}$^a$\color{black}, Deepak Jain\color{blue}$^b$\color{black},  Shobhit Mahajan{\color{blue}$^a$} , Amitabha Mukherjee{\color{blue}$^a$} and Nilza Pires\color{blue}$^c$ }

\def\I {\textnormal{\tiny\textsc{I}}}
\def\II {\textnormal{\tiny\textsc{II}}}
\def\III {\textnormal{\tiny\textsc{III}}}
\def\1 {\textnormal{\tiny\textsc{1}}}
\def\2 {\textnormal{\tiny\textsc{2}}}
\def\3 {\textnormal{\tiny\textsc{3}}}
\def\4 {\textnormal{\tiny\textsc{4}}}
\def\5 {\textnormal{\tiny\textsc{5}}}

\begin{document}
\maketitle
\small{ \color{blue}$^a$\color{black}Department of Physics $\&$ Astrophysics,  University of Delhi, New Delhi 110007, India\\
$~~~~~~~$ \color{blue}$^b$\color{black}Deen Dayal Upadhyaya College,
 University of Delhi, New Delhi 110015, India\\
$~~~~~~$ \color{blue}$^c$\color{black}Departamento de Física Teórica e Experimental, UFRN, Campus
Universitário, \\$~~~~~~~~~$Natal, RN
 59072-970, Brazil
}

$~~$E-mail: \textcolor{blue}{nrani@physics.du.ac.in}

\abstract{In this paper, we use the cosmokinematics approach to study the
accelerated expansion of the Universe. This is a model independent
approach and  depends only on the assumption that the Universe is
homogeneous and isotropic and is described by the FRW metric. We parametrize the deceleration parameter, $q(z)$, to constrain the
transition redshift ($z_t$) at which the expansion of the Universe
goes from a decelerating to an accelerating phase.
We use three different parametrizations of $q(z)$ namely,
$q_\I(z)=q_{\textnormal{\tiny\textsc{1}}}+q_{\textnormal{\tiny\textsc{2}}}z$, $q_\II (z) = q_\3 + q_\4 \ln (1 + z)$ and $q_\III(z)=\frac{1}{2}+\frac{q_{\textnormal{\tiny\textsc{5}}}}{(1+z)^2}$. A joint analysis of the age of galaxies, strong lensing and supernovae Ia data indicates that the transition redshift is less than unity i.e. $z_t<1$. We also use a nonparametric approach (LOESS+SIMEX) to constrain $z_t$. This too gives $z_t<1$ which is consistent with the value obtained by the parametric approach.}

\section{Introduction}
\label{sec:intro}
It is now widely  accepted that we are living in a Universe which is
accelerating at present \cite{aa1,ab2}.  Measurements of distance
moduli of Supernovae (SNe Ia) along with other independent
observations such as  results from the Planck satellite (Cosmic
Microwave Background (CMB), Baryonic Acoustic Oscillations (BAO) ),
etc. support this view \cite{ac3,ae5,af6}. However, the physics
behind this  late time cosmic acceleration  is still a
mystery. Cosmologists have  come up with different models to solve
this riddle \cite{ah8}. Further, the discovery of high redshift SNe Ia
provides strong evidence that the Universe was decelerating in the
past \cite{riess2007}. Therefore, it is very important  to find the
epoch of transition from the slowing down to  the speeding up of the
expansion of the Universe.

Earlier Shapiro and Turner (2005) used both parametric and
non-parametric methods to constrain the deceleration parameter, $q(z)$
and transition redshift, $z_t$  using SNe Ia data \cite{aj9}. Melchiorri et. al (2007)
also considered different  dynamical dark energy  models to put
constraints on $z_t$ \cite{bz28}. As expected, these constraints were
model dependent. Further in 2007, Rapetti et al. developed a new
kinematical method to study the dynamical history of the Universe
\cite{bz27}. By using the X-ray gas mass fraction measurements, they
supported the transition of the Universe from a decelerated to an
accelerated phase.

Many other authors have  used different data sets such as Gamma Ray
Bursts (GRBs), Hubble Observational Data $H(z)$,  BAO, CMB, Galaxy
Clusters , lookback time etc. to  reconstruct $q(z)$ or to constrain the transition
redshift \cite{ad4,bz29, ap16, ai10,al12, as19, avs, ats, ao15,
az26, aq17,ak11, ax28}.  Lima
et al. (2012) have   advocated many techniques that may constrain the
transition redshift effectively in the future \cite{an14}. In a
detailed study, Del Campo et al
proposed three different  $q(z)$ parametrizations  and put
constraints on the model  parameters by using SNe Ia, $H(z)$, BAO/CMB
data sets \cite{ax26}. Recently, Vargas dos Santos et al. (2015)
chose kink-like parametrizations of $q(z)$ and obtained
constraints on the free model parameters especially the duration of
the transition. They also used SNe Ia, BAO, CMB and {\it{H(z)}} data
\cite{dos}.

At present there is no well motivated theoretical model  of the
Universe which can explain all the aspects of this accelerated
expansion. Hence it is reasonable to  use a phenomenological
approach. In this paper we study a popular cosmographic approach to
characterize the properties of dark energy through the deceleration
parameter. Here we also attempt to
reconstruct  $q(z)$  through a parametric method.  This methodology
has both advantages and disadvantages. One of the plus points of this
technique is that it is independent of the matter-energy content of
the Universe. The only assumption in this approach is that the
Universe is  homogeneous and isotropic  on  cosmic scales and the FRW
metric is sufficient to describe the space-time of the Universe.  The
parametrization technique also helps to optimize future cosmological
surveys. On the other hand, this formulation does not explain the real
cause of the accelerated expansion. Further, the extracted value of
the present deceleration parameter may be dependent on the assumed
form of $q(z)$.

 In this work we use three different parametrizations of $q(z)$ to
 constrain the transition redshift in a model independent manner. We
 also use information criteria to compare these models. As mentioned
 above, earlier work based on $q(z)$ parametrizations used SNe,
 $H(z)$, BAO, CMB etc. to study
 either the model parameters or the transition redshift. {\it{In this paper
 we highlight the importance of strong  lensing data  to study the
 ``dynamic phase transition'' in the universe when it switched  from a
 decelerating  to an accelerating phase.  We further supplement this
 dataset with the age of galaxies and the recent data of SNe Ia
 (Joint Light Curve Analysis).}}

 To check the consistency of our results, we also constrain $z_t$ using a nonparametric approach. Many nonparametric approaches like Principal Component Analysis (PCA), Gaussian Process (GP), Non Parametric Smoothing method (NPS) etc have been used in the literature to constrain cosmological parameters \cite{pca}. Here, we use the LOcally wEighted Scatterplot Smoothing method (LOESS). This method locates a smooth curve among the data points without having any information about the priors i.e. the functional form estimation is done using raw data only. For details of the method see \cite{lo61,lo62,loess}. This method does not consider the contribution of the observational errors. Therefore, in order to include the effects of errors in the analysis, LOESS is combined with SIMEX (SIMulation EXtrapolation method) \cite{loess}.

The outline of the  paper is  as follows. In Section 2, we describe
the parametric approach to put constraints on deceleration parameter ($q(z)$) and transition redshift ($z_t$). In Section 3 we adopt a nonparametric approach to constrain $z_t$. Discussion is presented in Section 4. Finally, Section 5 contains some concluding remarks.\\


\section{Parametric Method}

\subsection{The Parametrizations of $q(z)$}
The reconstructed form of the equation of state of dark energy ($\omega_{\tiny{de}}(z)$) is usually written as \cite{qwn}
\begin{equation}
\label{eq1}
\omega_{\text{\tiny{de}}}(z)=\sum_{n=0}\omega_n x_n(z)
\end{equation}

By analogy, we write the parametrized form of the
deceleration parameter as
\begin{equation}
\label{eq2}
q(z)= \sum_{n=0}q_n x_n(z)
\end{equation}

For different $x_n(z)$ we get different parametrizations of $q(z)$:\\

Redshift:~~~~~~~~ $x_n(z)=z^n$

Logarithmic:~~~~$x_n(z)=[\textnormal{ln}(1+z)]^n$

Scale factor: ~~~~$x_n(z)=\left(\frac{1}{1+z}\right)^{2n}$\\

We use three different parametrizations of $q(z)$ corresponding to the above $x_n(z)$ \cite{ad4, aws, axs}.
\begin{equation}
\label{eq3}
q_\I(z)=q_{\textnormal{\tiny\textsc{1}}}+q_{\textnormal{\tiny\textsc{2}}} z
\end{equation}

\begin{equation}
\label{eq4}
q_\II (z) = q_{\textnormal{\tiny\textsc{3}}}+ q_{\textnormal{\tiny\textsc{4}}} \ln(1 + z)
\end{equation}

\begin{equation}
\label{eq5}
q_\III(z)=\frac{1}{2}+\frac{q_{\textnormal{\tiny\textsc{5}}}}{(1+z)^2}
\end{equation}

The first parametrization is the Taylor series expansion of
deceleration parameter around $z=0$ with
$q_{\textnormal{\tiny\textsc{1}}}$ as the present value and
$q_{\textnormal{\tiny\textsc{2}}}$ as the first derivative of $q(z)$
w.r.t. $z$. The second parametrization gives the present value of
the deceleration parameter as $q_{\textnormal{\tiny\textsc{3}}}$ at $z=0$
while at higher redshifts it diverges. The third parametrization
converges to $\frac{1}{2}$ as $z$ becomes large.

The Hubble parameter, $H(z)$ can be written in terms of $q(z)$ as

\begin{equation}
\label{eq6}
H(z)=H_0~\exp\left[\int\limits_0^z\frac{1+q(x)}{1+x}dx\right]
\end{equation}

where $H_0$ is the present value of the Hubble constant.

The corresponding Hubble parameter for the above mentioned parametrizations can be expressed as-

\begin{equation}
\label{eq7}
H_\I(z,p)=H_0(1+z)^{1+q_{\textnormal{\tiny\textsc{1}}}-q_{\textnormal{\tiny\textsc{2}}}}~\exp(q_{\textnormal{\tiny\textsc{2}}}z)
\end{equation}

\begin{equation}
\label{eq8}
H_\II(z,p)=H_0(1+z)^{1+q_{\textnormal{\tiny\textsc{3}}}}~\exp\left[\frac{q_{\textnormal{\tiny\textsc{4}}}}{2}[\ln(1+z)]^2 \right]
\end{equation}

\begin{equation}
\label{eq9}
H_\III(z,p)=H_0(1+z)^{\frac{3}{2}}~\exp\left[ \frac{q_{\textnormal{\tiny\textsc{5}}}(z^2+2z)}{2(1+z)^2}\right]
\end{equation}

Here $p$ represents the model parameters, i.e. the various $q_i's$.

\vspace{1cm}
\subsection{Datasets and Method}


\subsubsection{Age of galaxies}

We use two datasets of age of galaxies: Sample A and Sample B. Sample
A consists of the age of $32$ passively evolving galaxies in the
redshift range $0.117\leq z \leq 1.845$. This dataset is the
collection of three subsamples \cite{ar18}.

The first subsample contains 10 early type field galaxies. The SPEED
model is used to derive the age of these galaxies. It is assumed that
stars in an elliptical galaxy have a single metallicity value which is
an over-simplification. 
Further, the formation of stars in the single burst model may  bias
the age result as the formation of stars takes place at low redshift
also. This problem can be rectified by considering only the old
galaxies in each redshift bin. Jimenez et al. (2004) concluded from
their study that the systematic errors associated with age of galaxies
is not more than 10-15\% even with these assumptions
\cite{speed}.


The second subsample consists of 20 old passive galaxies released by
GDDS (Gemini Deep Deep Survey). The spectra obtained in this survey
are of very high quality. For this subsample, McCarthy et al computed
the synthetic spectra using the PEGASE.2 model and then compared the
obtained spectra with the observed spectral energy distribution
\cite{gdds}. Simon et al. further reanalyzed this subsample using the
SPEED model and obtained the age within 0.1 Gyr of the estimate of the
GDDS collaboration. Finally, they emphasized that systematics are not a
matter of concern for this subsample \cite{ar18}. Dust depletion may
also be a  cause of systematic error, but this effect is again
negligible.

The third subsample of sample A consists of two radio galaxies
53W091 and 53W069 \cite{nature, nolan}. While deriving the age for
these galaxies it was assumed that the large elliptical galaxy was
formed in a single burst after which there was no star formation in
it. This can cause over-estimation of age. However, the effect of
the metallicity on the calculated age was found to be negligible.
Following Samushia et al., we also use $12\%$ error in all the
observed age data points\cite{samu}.

We also work with another recent low redshift age dataset (Sample B)
which consists of LRGs (Luminous Red Galaxies) selected from the SDSS
DR7. This dataset consists of $12$ points in the redshift range
$0.05<z<0.4$ \cite{liu}. The full spectrum fitting method is used to
obtain the age of these galaxies by using the  single population
synthesis model (GalexEV/SteLib). The reason for using the combined
spectrum is to improve the signal to noise ratio and to reduce
contamination. Liu et al argue that the recent star formation in the
galactic center of massive galaxies may systematically underestimate
the obtained mean age at $z\leq 0.4$~ \cite{liu}. But this bias is not very
significant for galaxies with a large velocity dispersion. Following
this criterion, we also prefer to use subsample IV of the data (Liu G.,
private communication, 2015).

We use the minimum chi-square technique to find the best fit model parameters.
$$\chi^2=\sum_{i=1}^{n_1}\frac{[t^{th}(z_i,p)-\tau-t^{obs}(z_i)]^2}{\sigma_i^2}$$

where $n_1$ is the number of data points used in the sample, $t^{th}(z_i,p)$
is the theoretical age of the galaxy, $\tau$ is the delay factor or
incubation time, $t^{obs}$ is the measured current age of the galaxy and
$\sigma_i$ is the uncertainty in the observed age.

 The theoretical expressions for the ages of galaxies
 ($t^{\tiny{th}}(z_i,p)$) corresponding to the three parametrizations
 are as follows:

  \begin{equation}
  \label{eq10}
  t_\I(z,p)=H_0^{-1}\int\limits_0^{\frac{1}{1+z}}x^{q_{\textnormal{\tiny\textsc{1}}}-q_{\textnormal{\tiny\textsc{2}}}}~~\exp\left[-q_{\textnormal{\tiny\textsc{2}}}\left( 
    \frac{1-x}{x}\right) \right]dx
  \end{equation}

  \begin{equation}
  \label{eq11}
  t_\II(z,p)=H_0^{-1}\int\limits_0^{\frac{1}{1+z}}x^{q_{\textnormal{\tiny\textsc{3}}}}\exp\left[-\frac{q_{\textnormal{\tiny\textsc{4}}}}{2}\left[\ln~\left(\frac{1}{x}\right)\right]^2 
    \right]dx
  \end{equation}

  \begin{equation}
  \label{eq12}
  t_\III(z,p)=H_0^{-1}\int\limits_0^{{\frac{1}{1+z}}}x^{\frac{1}{2}}~\exp\left[- \frac{q_{\textnormal{\tiny\textsc{5}}}}{2}(1-x^2)\right]dx
  \end{equation}

  Here $p$ represents the model parameters, i.e. the various $q_i's$.

 In an earlier work, marginalization was done over the delay factor
 only and $H_0$ was kept fixed for the kinematic approach
 \cite{ad4}. The recent value of $H_0$ determined by SNe and Mega-masers and also through a combination of data including the age of old galaxies at intermediate redshift are $73.8\pm 2.4~ {\textnormal Kms^{-1}Mpc^{-1}}$ and $74.1\pm 2.2~ {\textnormal Kms^{-1}Mpc^{-1}}$ respectively \cite{apj1, apj2}. However, the Planck results suggest $H_0=67.8\pm 0.9~ {\textnormal Kms^{-1}Mpc^{-1}}$ \cite{rplank}. Therefore instead of choosing fixed value of $H_0$, we minimized $\chi^2$ w.r.t. $H_0$ in order to take care care of this discrepancy. We find that minimization and marginalization are
 equivalent in our case.
 \begin{equation}
 \label{eq13}
\chi^2=A+C\tau^2+2B\tau
 \end{equation}
 So we first minimize chi-square over the delay factor$(\tau)$. The
 minimum value of chi-square is
 \begin{equation}
 \label{eq14}
\tilde{\chi}^2=A-\frac{B^2}{C}
\end{equation}
  where

  $$A=\sum_{i=1}^{n_1}\frac{(t^{th}(z_i,p)-t^{obs}(z_i))^2}{\sigma_i^2},~~~~B=\sum_{i=1}^{n_1}\frac{(t^{th}(z_i,p)-t^{obs}(z_i))}{\sigma_i^2},~~~~
  C=\sum_{i=1}^{n_1}\frac{1}{\sigma_i^2}$$

 $\tilde{\chi}^2$ is again minimized over $H_0$
  \begin{equation}
  \label{eq15}
 \tilde{\tilde{\chi}}=\frac{d\tilde{\chi}^2}{dH_0}=\frac{dA}{dH_0}-\frac{2B}{C}\frac{dB}{dH_0}
 \end{equation}

 $$  \textnormal{where}~~~~~\frac{dA}{dH_0}=\frac{-2D}{H_0^3}+\frac{2G}{H_0^2}$$
 $$\textnormal{and} ~~~  \frac{dB}{dH_0}=-\frac{E}{H_0^2}$$

Putting the values of $ \frac{dA}{dH_0}$ and $\frac{dB}{dH_0}$  in Equation (\ref{eq15}) and  equating to $0$, we get
$$\frac{1}{H_0}=\frac{-JE+GC}{-E^2+CD}$$

where $$M=\sum_{i=1}^{n_1}\frac{[t_{obs}(z_i)]^2}{\sigma_i^2},~~~~J=\sum_{i=1}^{n_1}\frac{t_{obs}(z_i)}{\sigma_i^2},~~~~G=\sum_{i=1}^{n_1}\frac{\Delta(z_i,p)t_{obs}(z_i)}{\sigma_i^2}$$ 

  $$E=\sum_{i=1}^{n_1}\frac{\Delta(z_i,p)}{\sigma_i^2},~~~~D=\sum_{i=1}^{n_1}\frac{[\Delta(z_i,p)]^2}{\sigma_i^2},~~~~\Delta(z_i,p)=\frac{t^{\tiny{th}}(z_i,p)}{H_0^{-1}}$$

The final chi-square obtained after minimizing over both the delay
factor and the Hubble parameter is given by
    \begin{equation}
    \label{eq16}
  \tilde{\tilde{\chi}}_{age}^2=M-\frac{J^2}{C}-\frac{(GC-JE)^2}{C(E^2-CD)}
  \end{equation}

\subsubsection{Strong lensing}

When light passes near matter, it  bends due to the gravitational
field. The same phenomenon occurs if there is a galaxy or galaxy
cluster (lens) in the path of light coming from a bright object
(quasar) usually termed as the source.  If the source, lens and the
observer are placed in such a way that all three lie in the same line
of sight then a ring-like structure, called an Einstein ring, is formed
and the phenomenon is called strong lensing (SL). Two or more images can form in strong lensing (for details see \cite{at20}).

The phenomenon of strong lensing can be used to constrain cosmological
parameters. If we assume that the Singular Isothermal Sphere (SIS) or
Singular Isothermal Ellipsoid (SIE) model can be used to represent the
gravitational lens, then the radius of the Einstein ring is
\cite{au21, ax29, ast, asu}
\begin{equation}
\label{eq17}
\theta_E=4\pi\frac{\sigma^2_{\textnormal{\tiny\textsc{SIS}}}}{c^2}\frac{D_{ls}}{D_{os}}
\end{equation}

Here $\theta_E$ is the radius of the Einstein ring, $D_{ls}$ is the
angular diameter distance between the lens and the source, $D_{os}$ is
the angular diameter distance between the observer and the source and
$\sigma_{\textnormal{\tiny\textsc{SIS}}}$ is the velocity dispersion
of the mass distribution of the lens.

We define the ratio of the measured angular diameter distances, i.e.  $D_{ls}$ and $D_{os}$, as $\mathcal{D}^{obs}$:

\begin{equation}
\label{eq18}
\mathcal{D}^{obs} \equiv \frac{D_{ls}}{D_{os}}=\frac{c^2 \theta_E}{4\pi \sigma_{\textnormal{\tiny\textsc{SIS}}}^2}
\end{equation}

We select a subsample from the Sloan Lens ACS (SLACS) and Lens
Structure and Dynamics survey (LSD). This subsample consists of $70$ data
points in which lensing occurs due to galaxies while for the rest 10 lensing
occurs due to clusters. In treating lensing due to clusters, several assumptions are commonly made like the whole gas inside  is at same temperature, the gas pressure and gravity of the relaxed cluster balance each other according to hydrostatic equilibrium and the cluster is spherically symmetric. These assumptions may over-simplify the results and hence we do not include cluster data in our analysis. We also do not
consider the four-image lens systems as the SIS lens model produces
two images only. Finally we are  left with a subsample which consists
of $36$ data points. It is important to note that though the range
$0\leq \mathcal{D}^{obs} \leq 1$ is the only physically meaningful one
as $D_{ls}$ should always be smaller than $D_{os}$,  we have
also included points with $\mathcal{D}^{obs}>1$ since this is within 1$\sigma$. The redshift ranges of the lens and the source are
$0.106 \leq z_l \leq 1.004$ and $0.1965 \leq z_s \leq 3.9$
respectively (for details see \cite{au21}).

To constrain the best fit model parameter, chi-square is written as
\begin{equation}
\label{eq19}
\chi_{SL}^2(p)=\sum_{i=1}^{n_2}\frac{(\mathcal{D}_i^{th}(p)-\mathcal{D}_i^{obs})^2}{\sigma_{\mathcal{D}^{obs}}^2}
\end{equation}

where $n_2=36$ and $\sigma_{\mathcal{D}^{obs}}$ is the uncertainty in $\mathcal{D}^{obs}$.

The corresponding theoretical quantity $\mathcal{D}^{th}$ can be written as
\begin{equation}
\label{eq20}
\mathcal{D}^{th}(z_l, z_s;
p)=\frac{D_{ls}^{th}}{D_{os}^{th}}=\frac{D_A(z_l,z_s;p)}{
  D_A(0,z_s;p)}
\end{equation}

In flat Friedmann-Robertson-Walker (FRW) cosmology, the angular diameter distance is
\begin{equation}
\label{eq21}
D_A(z_l,z_s;p)=\frac{1}{(1+z_s)}\frac{c}{H_0}\int\limits_{z_l}^{z_s}\frac{dx}{E(x; p)}
\end{equation}

\begin{equation}
\label{eq22}
D_A(0,z_s;p)=\frac{1}{(1+z_s)}\frac{c}{H_0}\int\limits_{0}^{z_s}\frac{dx}{E(x; p)}
\end{equation}

\begin{equation}
\label{eq23}
 E(x;p)=\frac{H(x;p)}{H_0}
 \end{equation}
 In this method, the cosmological model enters through the ratio of
 the two angular diameter distances and so it is completely
 independent of the Hubble constant.

\subsubsection{Supernova Ia}

We use the latest Supernova Ia data which consists of $740$ data
points \cite{new1}. This includes data from low redshift ($z<0.1$),
SDSS-II ($0.05<z<0.4$) and SNLS 3 year ($0.2<z<1.0$) supernova
samples. It is assumed that the supernova will have the same intrinsic
luminosity at all redshifts provided they have identical colours, the
same galactic environment and shape. This assumption is expressed
through the  distance estimator which can be written as \cite{new2}
$$\mu_{obs}(\alpha, \beta; M)=m^*_B-M+\alpha~X_1-\beta~C$$
Here $m^*_B$ is the rest-frame B band peak magnitude, and $\alpha$,
$\beta$ and $M$ are the nuisance parameters. $X_1$ and $C$ describe the
stretching of the light curve and colour at maximum brightness
respectively.

The chi-square for the supernova analysis is written as

$$\chi^2_{\tiny{JLA}}(\mu_0, M; p_{th}, p_{obs} )=\sum_{i=1}^{n_3}\frac{[\mu_{th}(z_i; p_{th}, \mu_0)-\mu_{obs}(p_{obs};M)]^2}{\sigma_{\mu}^2}$$

where $n_3=740$, $p_{th}$ represents the model parameters and
$p_{obs}$ are the nuisance parameters present in $\mu_{obs}$,
i.e. $\alpha$ and $\beta$.
We fix the values of $\alpha$ and $\beta$ according to the
$\Lambda$CDM model as $0.14$ and $3.14$ respectively while we minimize
over the parameters $M$ and $H_0$ \cite{new2}.

Chi-square obtained after minimization over $M$ and $H_0$ is

$$\tilde\chi_{JLA}^2(p_{th},p_{obs})=A'-\frac{B'^2}{C'}$$

$$A'=\sum_{i=1}^{n_3}\frac{[\mu_{th}(z_i; p_{th})-\mu_{obs}(p_{obs})]^2}{\sigma_{\mu}^2}$$

$$B'=\sum_{i=1}^{n_3}\frac{[\mu_{th}(z_i; p_{th})-\mu_{obs}(p_{obs})]}{\sigma_{\mu}^2}$$

$$C'=\sum_{i=1}^{n_3}\frac{1}{\sigma_{\mu}^2}$$

$$\mu_{th}=5log_{10}(d_l)+\mu_0$$

where $d_l$ is the Hubble-free luminosity distance and

$$\mu_0 = 5log_{10}(c/H_0)+25$$

The combined chi-square is the sum of chi-squares obtained from all
three datasets i.e.

  \begin{equation}
  \label{eq24}
  \chi_{tot}^2= \tilde{\tilde{\chi}}_{age}^2+\chi^2_{\tiny{SL}}+\tilde
  \chi^2_{\tiny{JLA}}
  \end{equation}

\vskip 1cm

\subsection{Results}
We parameterize the deceleration parameter  with a two-parameter function ($ q_I(z), q_{II}(z)$)
and a one-parameter function ($q_{III}(z)$) to
understand  the expansion history of the Universe.. By fitting these functions
to the combined set of Age, SL and SNe Ia data, we
get the constraints described below.

\vskip 0.35cm
\textbullet \textbf{~~Parametrization I ~~$q_\I(z)= q_{\textnormal{\tiny\textsc{1}}}+q_{\textnormal{\tiny\textsc{2}}} z$}\\

\begin{figure}[ht]
  \centering
  \includegraphics[height=6.5cm,scale=6.5]{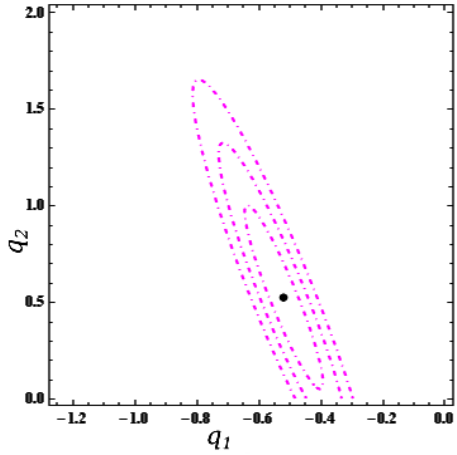}\\
    \caption{\small{1$\sigma$, 2$\sigma$ and 3$\sigma$ confidence
        level contours of chi-square in the
        $q_{\textnormal{\tiny\textsc{1}}}-q_{\textnormal{\tiny\textsc{2}}}
        $ plane for the combined datasets (Age+SL+JLA). The black dot
        shows the best fit value of the parameters.}}
        \label{fig1.1-nnlin}
\end{figure}

\begin{figure}[ht]
  \centering
  \includegraphics[height=4.5cm,scale=4.0]{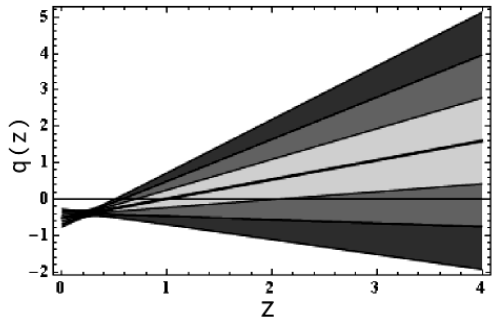}\\
    \caption{\small{The variation of $q_{I}(z)$ vs $z$ with 1$\sigma$,
        2$\sigma$ and 3$\sigma$ confidence regions. The thick black
        line is the best fit line. The horizontal line represents
        $q(z)=0$.}}
        \label{fig1.2-qqlin}
\end{figure}

Figure \ref{fig1.1-nnlin} shows $68.3\%$, $95.4\%$ and $99.7\%$ confidence
level ellipses in the model parameter plane for the combined
(Age+SL+JLA) dataset. Figure \ref{fig1.2-qqlin} shows the evolution of
$q(z)$ w.r.t. $z$ with $1 \sigma$, $2 \sigma$ and $3 \sigma$
errors.\\

The best fit parameter values for this parametrization are tabulated
in Table \ref{tab1}. The value of $z_t$ from the strong lensing data
is relatively large but combination of the three  datasets (Age+SL+JLA) lowers
it to a value which matches with other observations. The value of
$q_\I(0)$ for the combined analysis is $-0.52$.

\begin{table}[ht]
 \caption{\textbf{Values of best fit parameters for $q_\I(z)$} }
 \label{tab1}
 \centering
\begin{tabular}{|c|c|c|c|c|c|}
\hline \hline
Dateset & Age (Sample A) &Age (Sample B)  & SL &  JLA & Age(A+B)+SL+JLA  \\
 \hline
 $q_{\textnormal{\tiny\textsc{1}}}$ &-1.68 &-0.052 & -0.85 & -0.57 & -0.52\\
 \hline
 $q_{\textnormal{\tiny\textsc{2}}}$ &1.55 & 0.54 &  0.23 & 0.74 &0.53\\
 \hline
  $q_\I(0)$ &-1.68 & -0.052 &  -0.85 & -0.57 & -0.52\\
 \hline
 $\chi^2$ &  12.28 & 8.60 & 64.71 & 624.82 & 842.31 \\
 \hline
 $\chi^2_\nu$ & 0.41   &0.86 &   1.9 & 0.85 & 1.03\\
 \hline
 $z_t$ & 1.08 &0.10 &  3.70 & 0.77 & 0.98\\
 \hline
 FoM &  0.71& 0.007 &4.62 & 84.52 & 99.75\\
\hline
\end{tabular}
\end{table}

\newpage
\vskip 0.35cm
\textbullet ~\textbf{Parametrization II:~~$q_\II(z)=q_{\textnormal{\tiny\textsc{3}}} + q_{\textnormal{\tiny\textsc{4}}} \ln(1 + z)$}\\

\begin{figure}[ht]
  \centering
  \includegraphics[height=6.5cm,scale=6.5]{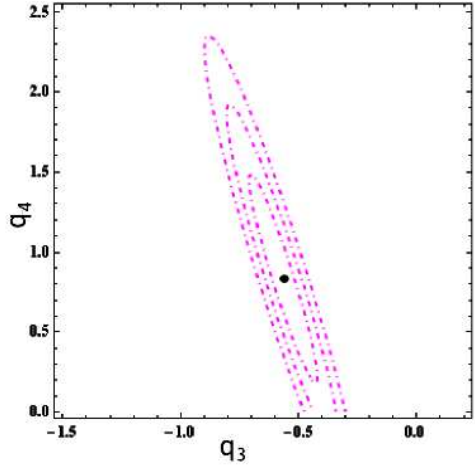}\\
    \caption{\small{$1\sigma$, $2\sigma$ and $3\sigma$ confidence
        level ellipse for the combined observations. Black dot
        represents the best fit value of the model parameters.}}
\label{fig2.1-nnlog}
\end{figure}

For the combined dataset, the 1$\sigma$, 2$\sigma$ and 3$\sigma$
confidence contours in the
$q_{\textnormal{\tiny\textsc{3}}}-q_{\textnormal{\tiny\textsc{4}}}$
plane are plotted in Figure \ref{fig2.1-nnlog}.
The value of $z_t$ obtained from the strong lensing data is once again on the  high
side, but by combining the  strong lensing, age and JLA data, we
get a value of $z_t$ which is consistent with other observations such
as $H(z)$ etc. within 3$\sigma$ level. The value of $q_\II(0)$ for the
combined dataset is $-0.56$. The best fit values of the model
parameters are given in Table \ref{tab2}.

\begin{table}[ht]
 \caption{\textbf{Values of best fit parameters for $q_\II(z)$ }}
 \label{tab2}
\centering
\begin{tabular}{|c|c|c|c|c|c|}
\hline \hline
  Dataset & Age (Sample A)&  Age (Sample B) &  SL &  JLA&Age(A+B)+SL+JLA\\
  \hline
  $q_{\textnormal{\tiny\textsc{3}}}$ & -1.88& -0.04 & -0.84 &-0.61&  -0.56\\
 \hline
 $q_{\textnormal{\tiny\textsc{4}}}$ & 2.54 &0.54& 0.30 &1.02&   0.83\\
 \hline
 $q_\II(0)$ &-1.88 & -0.04 &  -0.84 &-0.61&  -0.56\\
 \hline
 $\chi^2$ &12.35 &8.60&  64.76 & 624.49&841.38 \\
 \hline
 $\chi^2_\nu$ & 0.41 &0.86& 1.9&  0.85&1.03 \\
 \hline
 $z_t$ &1.09 &0.08&   15.44 &0.82&0.96\\
 \hline
  $FoM$ & 0.39 & 0.006 &  2.58 &65.32& 73.90 \\
 \hline
\end{tabular}
\end{table}

\begin{figure}[ht]
  \centering
  \includegraphics[height=5.5cm,scale=4.5]{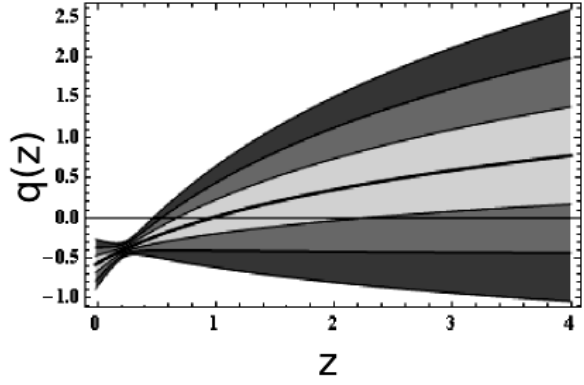}\\
    \caption{\small{Evolution of $q_{II}(z)$ with $z$. The filled
        regions are the 1$\sigma$, 2$\sigma$ and 3$\sigma$ confidence
        level regions respectively. The thick black line is the best fit
        line. The horizontal line represents $q(z)=0$.}}
        \label{fig2.2-qqlog}
\end{figure}

\newpage
\vskip 0.35cm
\textbullet~~\textbf{Parametrization III:~~$q_\III(z)=\frac{1}{2}+\frac{q_{\textnormal{\tiny\textsc{5}}}}{(1+z)^2}$}\\

\begin{figure}[ht]
  \centering
  \includegraphics[height=5cm,scale=4]{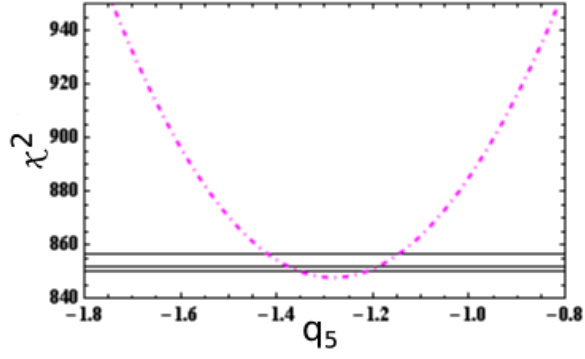}\\
    \caption{\small{Chi-square vs model parameter variation is shown
        by the Dot-Dashed line for the combined observations. Black
        horizontal lines are the 1$\sigma$, 2$\sigma$ and 3$\sigma$ lines.}}
        \label{fig3.1-nnp1}
\end{figure}

\begin{figure}[H]
  \centering
  \includegraphics[height=5.5cm,scale=4.5]{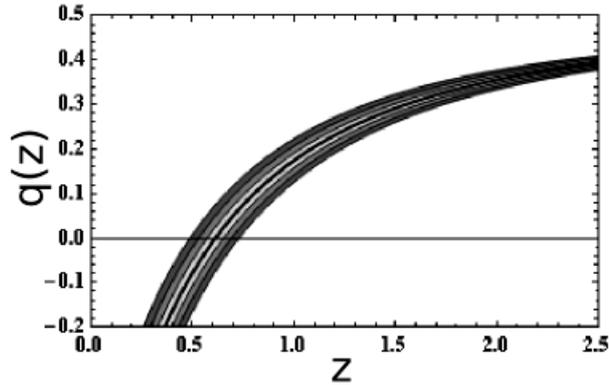}\\
    \caption{\small{The evolution of $q_{III}(z)$ with $z$. The  $1\sigma$,
        $2\sigma$ and $3\sigma$ confidence regions are shown by the
        filled regions. The thick black line is the best fit
        line. The horizontal line is the line corresponding to $q(z)=0$}}
        \label{fig3.2-qqp1}
\end{figure}

The value of best fit parameters for different datasets are tabulated
in Table \ref{tab3}. $q_\III(0)$ for the joint analysis is $-0.8$. Figures \ref{fig3.1-nnp1}  and  \ref{fig3.2-qqp1} show the variation of the chi-square with model parameters and the evolution of the deceleration parameter with redshift respectively.

\begin{table}[ht]
\caption{\textbf{Values of best fit parameters for $q_\III(z)$} }
\label{tab3}
\centering
\begin{tabular}{|c|c|c|c|c|c|c|}
\hline \hline

Dataset & Age (Sample A) &Age (Sample B) & SL & JLA& Age(A+B)+SL+JLA  \\
 \hline
 $q_{\textnormal{\tiny\textsc{5}}}$ &-2.58 & -0.57& -2.14 &   -1.3&-1.30\\
 \hline
 $q_\III(0)$&-2.08 & -0.07& -1.64&  -0.8&-0.80 \\
 \hline
 $\chi^2$ &12.47 & 8.67 &  66.13   &625.23&841.62 \\
 \hline
 $\chi^2_\nu$ &0.42& 0.86& 1.9&0.85   & 1.03\\
 \hline
 $z_t$ &1.27 &0.07& 1.07 &   0.61&0.60\\
 \hline
 \end{tabular}
\end{table}

\newpage
\subsection{Information Criteria}
Information criteria are used to rank different models that are available to describe a given dataset. We use the  Akaike Information
Criterion (AIC) and the Bayesian Information Criterion (BIC). For
details see \cite{ s1, s2, s3}

$$\textnormal{AIC}=\chi^2+2 P$$

$$\textnormal{BIC}=\chi^2+2 P \textnormal{ln} d$$

where $P$ and $d$ are the number of model parameters and data points
in the dataset respectively. $\chi^2$ is the minimum value of chi
square.
If $d/P<40$, it is good to use the corrected Akaike Information
Criterion ($\textnormal{AIC}_c$) \cite{s4}.

$$\textnormal{AIC}_c=\textnormal{AIC}+\frac{2P(P+1)}{(d-P-1)}$$
In our work $d/P>40$ and so we calculate AIC for all three
parametrizations. We also calculate BIC. To compare the models, we
calculate the differences given by $\Delta$AIC and $\Delta$BIC.
$$\Delta \textnormal{AIC}=\textnormal{AIC}_j-\textnormal{AIC}_{m}$$
$$\Delta \textnormal{BIC}=\textnormal{BIC}_j-\textnormal{BIC}_{m}$$

$\textnormal{AIC}_m$ and $\textnormal{BIC}_m$ are the minimum value of
AIC and BIC respectively. If more than one model fits the observations
equally, then the value of AIC and BIC will be minimum for the model
having the minimum number of model parameters.\\

\vskip 0.35cm
\textbullet~~\textbf{Summary of Information Criteria Results }\\

The results for different models using the  information criteria is
given in Table \ref{tab4}.
\begin{table}[ht]
\caption{\textbf{Information Criteria results for Age, SL and JLA data }}
\label{tab4}
\centering
\begin{tabular}{|c|c|c|c|}
\hline \hline

Model  & $P$ & $\Delta$ BIC & $\Delta \textnormal{AIC}$ \\
 \hline
 $q_{\tiny{I}}(z)$ & 2 & 14.11 & 4.70\\
 \hline
 $q_{\tiny{II}}(z)$ &2 & 13.18 & 3.77\\
 \hline
 $q_{\tiny{III}}(z)$ & 1 & 0.0 & 0.0 \\
 \hline
 \end{tabular}
\end{table}

$\bullet$ We also use the Figure of Merit (FoM) tool to quantify the  robustness
of the constraints obtained. For two parameters, FoM is defined as the
inverse of $95\%$ confidence limit area.  Tighter constraints in the
parametric space give large values of FoM's.\\

\section{Nonparametric Method}

\subsection{Hubble Data}

For this, we use recent Hubble data which includes $30$ data points. The redshifts of the passively evolving galaxies are known with high accuracy. So these galaxies can be used as chronometers and can provide the direct measurement of the $H(z)$. This is known as the differential age approach and 23 data points in this dataset are measured using this approach. The second way to measure $H(z)$ is by the clustering of the galaxies. In this method, $H(z)$ measurement is done by using the peak position of the BAO in the radial direction as the standard ruler. This approach is called clustering. Seven measurements of $H(z)$ are done by clustering. The complete dataset is taken from Yun Chen et al. \cite{asu}

\subsection{Method}
Non parametric method includes the study of the neighbourhood points of the focal point. For this we need to calculate the smoothing parameter or span $(s)$. This parameter tells us about the proportion of observations to be used in the local regression. $s$ ranges from $0$ to $1$. To find the appropriate value of $s$, we use the cross-validation method \cite{loess}. In this method, the cross-validation function $(CV)$ is defined as

\begin{equation}
CV(s)=\frac{1}{n}\sum_{i}(h(z_i)-\hat{h}(z_{-i}))^2
\end{equation}
where $n$ is the number of data points used in LOESS, $h(z_i)$ are the data points of the original data, $\hat{h}(z_{-i})$ are the fitted values obtained after omitting the $i^{th}$ observation from the local regression method at the point of interest i.e. $z_{i,0}$. $CV$ is calculated for different $s$ i.e. from 0.2 to 1. The value of $s$ which minimizes $CV$ is the best smoothing parameter for the particular data set. For the data set we use, $s=0.92$ is the best value.\\

The number of data points that we will use in LOESS can be calculated using the relation
\begin{equation}
n=N.s
\end{equation}

where $N$ is the total number of data points. In the data set we use, $N=30$ and $s=0.92$ so $n=27$.

The next  step of the method is the estimation of kernel in such a way that the points closer to the focal point ($z_{i,0}$) get more weight then the  farther points. This is because the points which are closer to the focal point are more correlated and so will affect the estimation of $\hat{h}(z_{i})$ more than the farther points. The weight or kernel function can be calculated using the following tricube function

\begin{equation}
w_{ij}= \left \{
\begin{array}{l l}
(1-u_{ij}^3)^3 & \quad {\textnormal{for}~~ |u_{ij}|<1}\\
0 & \quad \text{otherwise}
\end{array}
\right.
\end{equation}

where
\begin{equation}
u_{ij}=\frac{(z_j-z_{i,0})}{\Delta}
\end{equation}

$\Delta$ is the maximum difference between the focal point and the last ($j^{th}$) element of its window i.e. max$|z_j-z_{i,0}|$.\\

To calculate the value of reconstructed $h(z)$ i.e. $\hat{h}(z_i)$, we define chi-square as
\begin{equation}
\chi^2=\sum_{i=1}^n w_{ij}(h(z_i)-a-(b*z_i))^2
\end{equation}

The value of $[a+(b*z_i)]$ which minimizes chi-square gives the value of $\hat{h}(z_i)$. This process is repeated for all data points to find the reconstructed value corresponding to all redshifts given in the real data.\\








\subsection{Results}

\begin{figure}[ht]
  \centering
\includegraphics[height=5.5cm,scale=4]{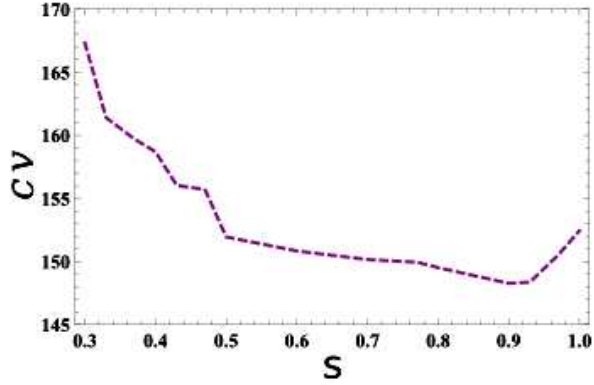}\\
    \caption{\small{CV vs s plot for the Hubble data}}
\label{cvvv}
\end{figure}

The $CV$ versus $s$ plot is shown in Figure \ref{cvvv}. This shows that the smoothing parameter value which minimizes the cross validation function is $0.92$.\\

\begin{figure}[ht]
  \centering
\includegraphics[height=6cm,scale=5.5]{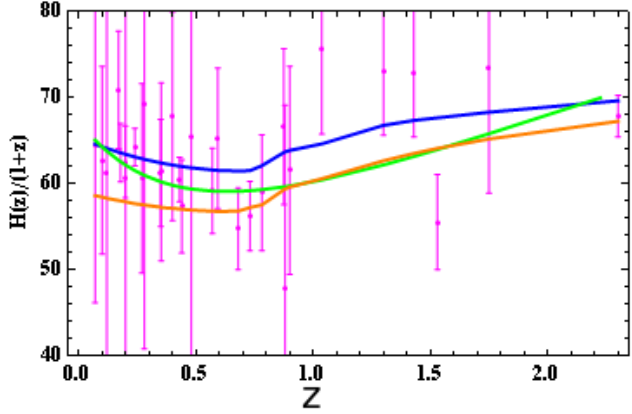}\\
    \caption{\small{Variation of $H(z)/(1+z)$ with $z$; the blue line corresponds to the reconstructed values through LOESS, the orange line shows reconstructed values from LOESS+SIMEX and the green line is for $\Lambda$CDM} model.}
\label{hzbyz}
\end{figure}

Figure \ref{hzbyz} shows the variation of $H(z)/(1+z)$ with $z$. The scattered points are the values of $H(z)/(1+z)$ from the data with error bars.The blue line represents the reconstructed values of $H(z)/(1+z)$ using the LOESS method. In the LOESS method observational errors are not taken into account. This can be done using the SIMEX. The results of using LOESS with SIMEX is shown by the orange line in Figure \ref{hzbyz}. The detail of SIMEX are given in \cite{loess}. It is clear from the plot that the transition from decelerated expansion to accelerated expansion occurs at $z_t\sim 0.7$. For reference we plot the $H(z)/(1+z)$ for $\Lambda$CDM model (green line) which gives $z_t \sim 0.6$.

\vskip 1cm
\section{Discussion}

In the present analysis  we use a sample of age data (passively evolving
galaxies+LRGs) which has $44$ points and is therefore larger than the data used earlier.
As pointed out earlier, Jimenez et al. (2004) showed that the  uncertainties
associated with the  age data may not be more than $10-15\%$
\cite{speed}. Recently, Jun Jie Wei et al. analysed the data with $24\%$
uncertainty in the observed
age of passively evolving galaxies in order to take care of some of
the unknown systematics which are not sufficiently understood \cite{dis2}. As
expected, the constraints obtained on the model parameters were
weaker. Wei et al. also relaxed the
assumption of a uniform delay factor for all galaxies. Instead they
considered the case in which the delay factor is distributed through
a Gaussian and a Top Hat distribution. They conclude that the delay factor
distribution does not affect the result significantly. In the present
work, {\it{we  minimize both the effect of delay factor and Hubble
    parameter analytically, something we believe has  not been  addressed completely in the
    past.} }\\


Strong gravitational lensing is another important tool used in
this work to constrain the transition redshift. Recent work done by
different groups has made this probe more popular to constrain the
cosmological parameters \cite{au21,ax33,ax31,nxt}. But there are many
systematic issues with this method. The three important
uncertainties associated with this method are the unknown mass
distribution of the galaxies, stellar velocity dispersion and the line
of sight mass contamination. Further, the data for lensing is drawn
from different surveys which may also cause a systematic error.\\

As the true mass profile of the galaxies is not known, various models
have been proposed in the literature. But due to the simplicity, most
frequently used  are the SIS (Singular Isothermal Sphere) and SIE
(Singular Isothermal Ellipsoid) models. The advantage of considering
the SIS model
is that, to a first approximation, it describes the average properties
of the galaxies quiet
accurately. However, real galaxies
show some deviation from the assumed symmetry. Models including
ellipticity provide a better description of mass distribution but they
need a comparatively  large number of parameters.\\

As mentioned earlier, the total density distribution of the lenses is
described by the SIS model (stellar and dark matter halo). In
principle, the velocity dispersion of the stellar component might be different
from that of the dark component. This fact is supported by X-ray properties of
late elliptical galaxies. Treu et al. (2006), after analysing the
large sample of lenses of SLACS
survey \cite{av22}, noticed that the average
ratio of $\sigma_0$ and $\sigma_{SIS}$ (for the total mass) is very
close to unity. This result suggests that some unknown nexus exists
between the stellar and dark matter halo which is sometime referred to
as the "bulge
halo conspiracy". In practice, $\sigma_0$ and $\sigma_{SIS}$ are
related as

$$\sigma_{SIS}=f_{\tiny{E}} \sigma_0$$

Here $f_E$ is a free parameter
which incorporates  the uncertainties like
 the error introduced in $\theta_E$ due to the SIS
assumption, the rms error
 between the observed velocity dispersion and model velocity
 dispersion etc. In literature, $f_E$ has been treated in a variety of
 ways. Melia et al. for instance, kept it as a free parameter
 \cite{ax31}. Wang and Xu present a new method which completely
 eliminates the uncertainty appearing due to $f_E$
 \cite{ax32}.
Van de Ven et al have   found that  $\sqrt{0.8} < f_E < \sqrt{1.2}$
\cite{van}. In this work we assume $ f_E = 1$.\\


The third systematic is due to line of sight mass contamination
\cite{ax33, ax34, ax37}. In the present work, lenses are assumed to be
isolated.
In principle the line of sight density fluctuations can affect the
lens model also. Bar-kana showed that large scale structure (LSS)
changes the angular diameter distance at a given redshift
\cite{ax35}. This shows that if the effect of LSS, is not included while
reconstructing the lens, it may lead to incorrect conclusions. This
effect dominates when the  source and the  lens are at large
redshifts. Keeton et
al claim that external shear perturbation is small due to LSS so
its  effect on gravitational lenses can be neglected \cite{ax36}. A
larger dataset for  strong lensing  may help to get rid of this
effect.\\

The total uncertainty present in $\mathcal{D}^{obs}(\sigma_0, f_E,
\theta_E)$ is calculated by using the propagation equation.
The uncertainty both in the Einstein radius $\theta_E$ and
the velocity dispersion $\sigma_0$ are $\sim5\%$ each \cite{aw23}. The
error in  $f_E$ is nearly $6\%$ \cite{ ax31,av22,van}. So the total
error in $\mathcal{D}^{obs}$ comes out to be approximately $16\%$ .\\

\vskip 0.5cm
\section{Conclusions}

More than four  decades ago,  Allan Sandage (1970) argued that
we need a precise measurement of the  Hubble constant ($H_0$) and
the deceleration parameter $q(z)$ to distinguish between various
cosmological models \cite{aa22}. But with the progress in the
cosmological observations, it is believed that we may require more
than two numbers to understand  the expansion history of the
universe \cite{apj}. In this regard, the
reconstruction of  $q(z)$ and the study of the transition redshift are
very important in describing the evolution history of the universe.\\

One of the objective of this work is to include strong lensing (SL) data for the reconstruction of $q(z)$ through a  parametric approach.
There are many advantages of using SL data in the form of distance ratio.
Firstly, this ratio is independent
of the Hubble parameter. Secondly, this data can lift the degeneracies in the determination of the constraints obtained from the
other observations. Finally, this method does not depend upon the
evolutionary effect  of the sources and is also immune to dust absorption.
In contrast
to the SL data,  the constraints obtained from the age data  depend upon the
Hubble parameter. Therefore we have  minimized the $\chi^2$
w.r.t. $H_0$ and the delay factor analytically, which was not addressed
completely in the past \cite{ad4}.\\

The constraints obtained on the transition redshift by
using strong lensing, age of the  galaxies and recent SNe Ia
data through a {\it{parametric approach}} are as follows:

$1.$ All the three $q(z)$ parametrizations are consistent with an
accelerating universe at the present epoch and the best fit value of
$z_t$ is less than $1$.

$2.$ Within $3 \sigma$ confidence level, both the present values
of the deceleration parameter and  the transition redshift are in
concordance with the constraints obtained from the other observations
such as {\it{H(z)}}, BAO, Clusters etc.

$3.$ The evolution of the deceleration parameter w.r.t. the redshift
shows that there is no slowing down of the  present cosmic
acceleration as reported earlier \cite{1306, 1407}.

$4.$ We observe the tension between the results obtained from the
high redshift ( Sample A) and the low redshift age data ( Sample
B). This inconsistency disappears when it is combined with other
datasets.

$5.$ The present value of $q(z)$ and $z_t$ obtained from these
parametrizations  ($q_{I}, q_{II}$)  have large error bars at high
redshift. This may be
due to lack of data points and more importantly the uncertainties
 associated with the lensing data set. The present strong lensing data
 is not precise enough to constrain the model parameters tightly. Therefore, we use the combination of age, strong lensing and SNe Ia data. These observations have different restrictive powers in the parametric space. Hence, it is always better to use complementary tools to put constraints on the model parameter. This also helps to break the degeneracy inherent in the parametric space. With the addition of age data and SNe Ia data, the constraints on the
parameters improve and become consistent with the $\Lambda$CDM model.

$6.$ We use two information criteria (AIC \& BIC) for selecting
 the  best parameterization.
 $\Delta$BIC greater than $6$ is considered
 as strong
evidence against the model with the higher BIC \cite{s4}. The rule remains
same for $\Delta$AIC. In our calculation, $\Delta$AIC and $\Delta$BIC
are  minimum for the third parametrization as expected. The value of
$\Delta$BIC is greater than $6$ for the  first and second parametrizations,
which indicates  a positive evidence against these models. We believe
that the IC analysis
favours the third parametrization.

In order to obtain the effective  constraints using SL data, one must
control the associated systematics. One very important
parameter linked with SL measurements is the observed velocity
dispersion (${D_{ls}/D_{os}} \propto \sigma^{2}$). As reported by Schwab
et al. \cite{ns}, the  actual velocity dispersion is luminosity
weighted along the
line of sight and also over the spectrometer aperture. In addition to
all these factors, one should also take care of the radial and tangential
component of velocity dispersion.  After including the above factors in
the analysis,   $\mathcal{D}^{obs}$ is
 replaced by \cite{syksy}

 $$N(\alpha, \beta,
\gamma, \delta) {\theta_{E}^{\alpha - 1} \over {4\pi \sigma_0^2}}$$

This modified ratio takes care  of the anisotropy  present in the velocity
 dispersion as well as other lens parameters.

As $\mathcal{D}^{obs}$ is dependent on $\sigma_0, N(\alpha, \beta,
\gamma, \delta)$ and $\theta_E$, it is crucial to check the sensitivity of these parameters on the
$\mathcal{D}^{obs}$. If we substitute the mean values of these parameters, we find that the value of $\mathcal{D}^{obs}$ gets reduced by around $12 \%$. What is more surprising is that this new data set of reduced $\mathcal{D}^{obs}$  leads to a universe which is decelerating at present, something which is not supported by observations. This leads us to conclude that our understanding of the lens quantities needs to improve further.

The point we want to highlight is that the determination of lens
parameters as well as the observed quantities in the SL data are very
crucial. Small changes in the lens parameters may change the result
completely. Therefore a  better understanding of the assumptions and
the systematics associated with the lens parameters must be addressed.

The  obtained  constraints are
limited by the quantity and  the quality of the data for strong
lensing. Nevertheless, the obtained results are meant to highlight the
importance of strong lensing as well as the updated age data which
gives an  independent estimate of the cosmological parameters.
 At present the known gravitational lenses are of the order $\sim
 10^{2}$. It is expected that next generation space telescopes such as
Euclid will observe $\sim 10^5$ strong galaxy-galaxy gravitational
 lenses. So in the near future, we can expect that  the enhanced  data
 sets will allow us to put tighter
 constraints on the cosmological parameters \cite{sergent}.
Similarly it is also expected that future surveys such as BOSS will
observe a large number of quiescent massive LRGs and hence  more accurate
age-redshift data of galaxies will be obtained \cite{boss}. This may
further reduce the systematic uncertainties associated with age data.\\

The constraints obtained from the {\it{nonparametric}} approach are summarized as follows:

$7.$ We have applied a nonparametric method (LOESS) to constrain $z_t$ by using the Hubble data only because this technique can reconstruct the global trend of the observed quantities without assuming any prior or cosmological model. So by studying the variation of $H(z)/(1+z)$ w.r.t. $z$ through LOESS method one can constrain $z_t$ easily. This method gives a smooth curve using a nonparametric regression through the real scattered data points. The value of $z_t$ comes out to be about $0.7$. This is in agreement with the value we obtained using the model independent (cosmokinematics) approach. Further, the transition redshift obtained from the LOESS+SIMEX analysis does not change significantly but it scales down the variation of $H(z)/(1+z)$ as compared to its values obtained from the LOESS only. But we did not use Hubble data in the parametric approach as this dataset has been used recently for constraining the deceleration parameter \cite{dos}. 

 The most important point is that  knowledge of the nature of the dark sector of the
Universe is still woefully inadequate. We believe therefore that we  should use model independent
techniques (both parametric  as well as nonparametric) with  all available datasets for extracting as much information as possible about the Universe and its structure.\\

\textbf{Acknowledgments}

Authors are thankful to the anonymous referee for useful comments. NR acknowledges financial support from the UGC Non-NET scheme
(Govt. of India) and the facilities provided at IUCAA Resource Centre,
Delhi University. NR also thanks Remya Nair (IUCAA, Pune) and
Sanjeev Kumar (DU) for helpful discussions. One of the author (DJ) is
thankful to CTP (Jamia Milia Islamia, New Delhi) for research support. AM thanks Research Council, University of Delhi, Delhi for providing support under R \& D scheme 2014-15.
 Authors are greatful to Lixin Xu, Fulvio Melia, Marek Biesiada, Ariadna Montiel and Vincenzo Salzano for useful suggestions and discussion.



\end{document}